**Full title:** Reliability for Nerve Fiber Layer Reflectance Using Spectral Domain Optical Coherence Tomography

**Short title:** NFL Reflectance Analysis for Reliability testing

**Word count:** 5053

**Tables:** 6; **Figures:** 2

**Section code:** GL

**Author:** Kabir Hossain, PhD, Ou Tan, PhD, Po-Han Yeh, Jie Wang, Elizabeth White, Dongseok Choi, David Huang

Casey Eye Institute, Oregon Health & Science University

***Corresponding Author**: Ou Tan

Associate Professor of Research Ophthalmology

Casey Eye Institute

Oregon Health & Science University

5034949436

Email: tano@ohsu.edu

**Funding Sources:** NIH grants R01 EY023285, R21 EY032146 and P30 EY010572, and an unrestricted grant from Research to Prevent Blindness to Casey Eye Institute.

**Commercial relationships:**

OT: P; DC: None; AC: None; DH: F,R,I,P, Others: None

# Abstract

**Purpose**: Reliability for Nerve Fiber Layer Reflectance Using Spectral Domain Optical Coherence Tomography (OCT)

**Methods**: The study utilized OCT to scan participants with a cubic 6x6 mm disc scan. NFL reflectance were normalized by the average of bands below NFL and summarized. We selected several reference bands, including the pigment epithelium complex (PPEC), the band between NFL and Bruch's membrane (Post-NFL), and the top 50% of pixels with higher values were selected from the Post-NFL band by Post-NFL-Bright. Especially, we also included NFL attenuation coefficient (AC), which was equivalent to NFL reflectance normalized by all pixels below NFL. An experiment was designed to test the NFL reflectance against different levels of attenuation using neutral density filter (NDF). We also evaluated the within-visit and between-visit repeatability using a clinical dataset with normal and glaucoma eyes.

**Results**: The experiment enrolled 20 healthy participants. The clinical dataset selected 22 normal and 55 glaucoma eyes with at least two visits form functional and structural OCT (FSOCT) study. The experiment showed that NFL reflectance normalized PPEC Max and Post-NFL-Bright had lowest dependence, slope=-0.77 and -1.34 dB/optical density on NDF levels, respectively. The clinical data showed that the NFL reflectance metrics normalized by Post-NFL-Bright or Post-NFL-Mean metrics had a trend of better repeatability and reproducibility than others, but the trend was not significant. All metrics demonstrated similar diagnostic accuracy (0.82-0.87), but Post-NFL-Bright provide the best result.

**Conclusions**: The NFL reflectance normalized by the maximum in PPEC had less dependence of the global attenuation followed by Post-NFL-Bright, PPEC/Mean, Post-NFL-Mean and NFL/AC. But NFL reflectance normalized by Post-NFL-Bright had better result in two datasets.

# 1 Introduction

Retinal Nerve Fiber Layer (RNFL) thickness, phase retardation, and reflectance are important bio-markers for glaucoma diagnostics [1]. Especially NFL thickness by Optical Coherence Tomography (OCT) has been a well-established method to monitor glaucoma progression and confirmation. However, for mass screening, NFL thickness is not suitable to be used alone [2]. It could be noted, Retinal ganglion cells (RGC) and their axons are significantly damaged by glaucoma which leads to visual field abnormalities and vision loss [1,3]. A study shows that RNFL reflectance is more sensitive to glaucoma damage than a change in RNFL thickness [3]. Further, a decrease in RNFL reflectance occurs before thinning of the RNFL [6]. Hence, an investigation into the RNFL reflectance analysis technique is imperative.

NFL reflectance was derived from OCT image is sensitive to different attenuations due to the optical scattering properties, e.g., media opacity, poor focusing, and ocular media opacities. These effects are often compensated by reference layers [1,4,5]. In [1] and [4], the RNFL reflectivity is normalized by the Retinal Pigment Epithelium (RPE). In [1] and [4], the RNFL reflectivity is normalized by the Retinal Pigment Epithelium (RPE). However, Gardiner SK at el. [5] showed that NFL reflectance corrected by the Post-NFL band had better repeatability than the RPC band. It could be noted, upon observation, we noticed that the Post-NFL band has a darker layer, which could be a disadvantage. However, by selecting the top 50% of pixels, we can opt out of those regions and potentially gain better advantages. Therefore, we selected the top 50% of pixels from the Post-NFL band and named it Post-NFL-Bright. Moreover, we observed that more layers below the NFL provide better repeatability; hence we considered the summation of attenuation coefficients (AC). The attenuation coefficient is an optical property that explains the attenuation of light that occurs due to the scattering and absorption properties of tissue. Therefore, the determination of the attenuation coefficient provides valuable information on glaucoma. Study [3] shows that the average RNFL attenuation coefficients are fully separable for normal and glaucoma. Several papers [4, 6,7,8] have been published for the quantitative analysis of attenuation coefficients. In this depth-resolved [7] based technique was used to determine RNFL attenuation coefficient maps based on a method that uses the retinal pigment epithelium as a reference layer. We have extended the depth-resolved based method by summation of the attenuation coefficient (AC) in NFL layers (which we called NFL/AC) because we want to compare and evaluate the NFL/AC together with other references (NFL/PPEC, NFL/Post-NFL-Mean and NFL/Post-NFL-Bright). Another reason is that, as mentioned earlier, we observed that more layers provide better repeatability, and the AC used more layers. The primary goal is to identify a metric that not only provides better repeatability and reproducibility but is also reliable across all datasets.

To evaluate the metrics, this study utilized two cohorts of datasets: Neutral Density Filter (NDF) experimental and the Functional and Structural OCT (FSOCT). In the NDF experiment, a set of seven optical density levels were employed, consisting of NDF with optical densities ranging from 0.1 to 0.6, in addition to NDF without any filter. It could be noted here cataract not only affect the peripapillary RNFL thickness but also signal strength [9]. The reasons we used NDF experimental data in the analysis because we wanted to test dependency of the normalized NFL reflectance in each NDF levels. The FSOCT datasets divided on two sets based on number of scans for within-visit repeatability and between visit reproducibility.

In the final analysis, we evaluated the references to test the dependencies of NFL reflectance on NDF levels, as well as the repeatability, reproducibility, and diagnostic accuracy of the FSOCT dataset. We estimated the slope against each NDF level with the NDF dataset and calculated the pooled standard deviation (pooled SD) and AROC for within-visit repeatability, between-visit reproducibility, and diagnostic accuracy, respectively.

## 2 Methods

### 2.1 Participants

This study utilized two cohorts of datasets, Neutral Density Filter (NDF) experimental data and Functional and Structural OCT (FSOCT) data. This prospective observational study was performed from January 06, 2017 to May 30, 2019 for FSOCT and 3/8/2022 to 5/12/2022 for NDF experiment at the Casey Eye Institute, Oregon Health & Science University (OHSU), Portland, OR, USA. The Institutional Review Board at OHSU approved the research protocol to carry the research accordingly with the tenets of the Declaration of Helsinki. Each of the participant given their written informed consent. This research was performed according to the Health Insurance Portability and Accountability Act of 1996 (HIPAA) privacy and security regulations.

The participants in the NDF study were part of the "Pilot Studies for New Scan Protocols Using Ultrahigh-Speed Optical Coherence Tomography" study. The NDF data contains only a normal group, while the FSOCT data consists of both a normal and a glaucoma group.

The inclusion and exclusion criteria for both the NDF experimental data and the FSOCT normal group are as follows: (1) No evidence of retinal pathology or glaucoma, (2) a normal Humphrey 24-2 VF, (3) intraocular pressure < 21 mm Hg, (4) central corneal pachymetry > 500 μm, (5) no chronic ocular or systemic corticosteroid use, (6) an open angle on gonioscopy, (7) a normal appearing optic nerve head (ONH) and NFL, and (8) a symmetric ONH between left and right eyes.

It could be noted that the participants were excluded from this study if any of the following situations were observed: (1) best-corrected visual acuity less than 20/40, (2) age < 40 or >80 years for FSOCT and age>21 for NDF, (3) spherical equivalent refractive error of > +3.00D or < -7.00 diopters, (4) previous intraocular surgery except for an uncomplicated cataract extraction with posterior chamber intraocular lens implantation, (5) any other diseases that might cause VF loss or optic disc abnormalities, or (6) inability to perform reliably on automated VF testing.

The FSOCT glaucoma group further divided into two groups, Preperimetric glaucoma (PPG) and Perimetric glaucoma (PG). The inclusion criteria for each of the group were different. The PG group's inclusion criteria were (1) an optic disc rim defect (thinning or notching) or retinal NFL defect visible on slit-lamp bio-microscopy, and (2) a consistent glaucomatous pattern on both qualifying Humphrey SITA 24-2 VFs. A glaucoma specialist assed the pattern of glaucoma defect based on the VF total deviation map. It could be noted that either pattern standard deviation (PSD) outside normal limits ($p < 0.05$) or glaucoma hemifield test outside normal limits were defined as the abnormality criteria for glaucomatous VF. The PPG group only meet the biomicroscopic criteria (1), but not the VF criteria (2).

It should be noted that FSOCT normal data were followed up every year, while the glaucoma data were followed up every six months.

## 2.2 Data Acquisition

A spectral-domain OCT system (Solix, Optovue, Inc., Fremont, CA, USA) was used for scanning participants with a 130 kHz scanning rate, 840 nm wavelength. The study utilized the optic disc volumetric high-definition OCT angiography (HD OCTA) scan.

The area covered by the optic disc volumetric HD OCTA scan was 6.0x6.0 mm that centered on the disc. The cross-sectional B-frames consisted of 512 A-lines. Each cross-sectional B-frames were repeated twice at each location to allow the computation of the angiographic flow signal, and each volume composed of 512 B-frames. An orthogonal registration algorithm was used to merge two volumetric scans, i.e., a vertical-priority raster and a horizontal-priority raster, to reduce motion artifacts and improved image quality. The resultant merge algorithm provided both angiographic (flow signal) and structural (reflectance signal) images. In the study, we used good quality images with a signal strength index (SSI) of 50 (out of 100) or more and a quality index (QI) of 5 (out of 10) for FSOCT dataset. For NDF, we didn't use SSI or QI, since want to study NDF experimental data with no quality control.

Standard automated perimetry on the Humphrey Field Analyzer (HFA II; Carl Zeiss Meditec, Inc., Dublin, CA, USA) used to assess the VF, which is using the Swedish Interactive Thresholding Algorithm 24-2.

## 2.3 NFL Reflectance Analysis

### 2.3.1 Nerve Fibre layer (NFL) Reflectance Metrics

We have evaluated several reference bands in this study (See Figure. 1), such as pigment epithelium complex (PPEC), the band between NFL and Bruch's membrane (Post-NFL), and created a new reference called Post-NFL-Bright by selecting the top 50% of pixels with higher values from the Post-NFL band. Additionally, summation of attenuation coefficient in NFL layers, NFL/AC also evaluated for reliability test. A custom MATLAB program has been used to analyzed the NFL reflectance, the OCT data were transferred into linear scale first and then converted in to dB scale later. It could be noted the blood vessels were excluded from the OCT data for estimation. The NFL reflectance were summed to create NFL reflection map, then the ratio for each of the metrics (NFL/PPEC Max, NFL/PPEC mean, NFL/Post-NFL-Mean and NFL/Post-NFL-Bright) were estimated. Detailed description of the references can be found here [2].

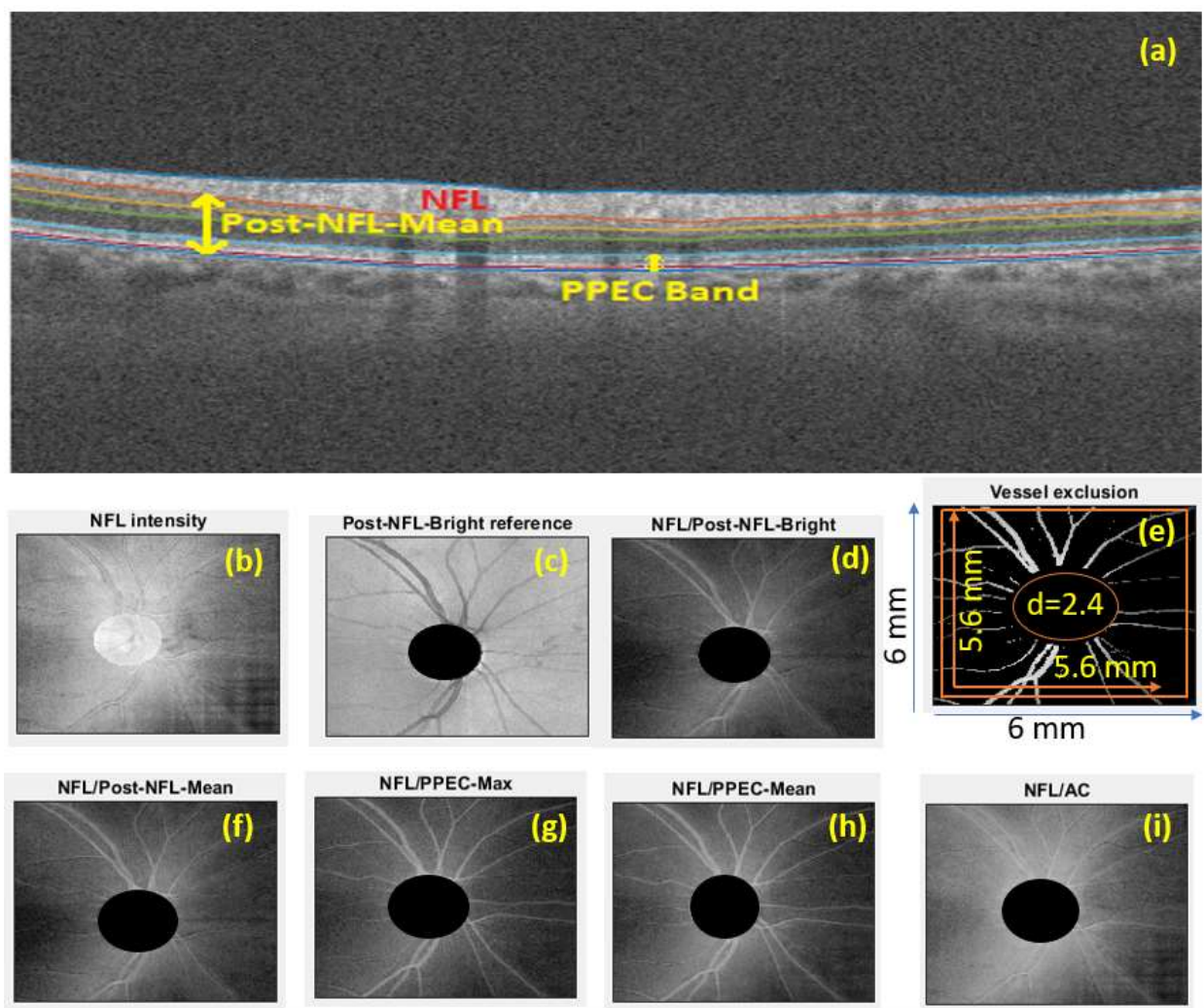

**Figure 1.** Illustration of reference and nerve fiber layer (NFL) reflectance map (A) NFL is defined as the band between ILM and post NFL boundary; Post-NFL is defined the bands between post-NFL boundary and BM; PPED is defined as the bands between EZ and BM. (B) NFL intensity is the summation of NFL reflectivity in the NFL band. (C) A reference is the summation of the reflectivity in the reference bands. Here the example is form post-NFL-Bright reference band, which is defined as the bright bands in the post-NFL bands, or the top half if we sort the OCT pixels in the post-NFL bands based on intensity. (D) NFL reflectance normalized by reference post-NFL-Bright. (E) Disc and larger vessel was detected. The overall average of NFL reflectance is only on the mask without disc and vessels. (F)-(I) NFL reflectance normalized by other reference: Post-NFL-Mean, PPEC-Max, PPEC-Mean and attenuation coefficient (AC).

### 2.3.2 Summation of Attenuation Coefficients (AC)

A study suggested that the glaucomatous damage is not only reflected by the thickness and reflectance of the RNFL [6] but also by changes in its optical scattering properties. Therefore, quantitative scattering property from the OCT images is needed to measure. To measure the summation of the attenuation coefficient, we estimate the ratio of OCT intensity by the summation of intensity multiply by pixel size. We have implemented the Depth-resolved model-based reconstruction [7] method to estimate the summation of the attenuation coefficient. It could be noted that the author [7] used average attenuation coefficients whereas we used the summation for AC estimation because we want to compared the AC with references.

### 2.3.3 NDF Experiment

A series of scans for neutral density filter (NDF) were conducted for each scan pattern. Baseline scans were taken without any NDF. The NDF scans were taken by placing an absorptive NDF of increasing optical density (NEK01; Thorlab, Newton, NJ, USA) in front of the eye. The optical densities ranged from 0.1 to 0.6, as well as NDF with no filter. This resulted in the collection of 7 scans for each participant.

Later, the NDF data were used to normalize the NDF reflectance with the references. We applied a linear mixed model to estimate the slope for each metric at each NDF level. Lower slope values for the metrics indicate effective compensation in removing artifacts, whereas higher slope values indicate less effective compensation. Moreover, to assess the difference in slope between two segments, we utilized a broken stick model that had a breakpoint at NDF level = 0.4 optical density.

### 2.3.4 Adjustment of confounding factor

Our prime objective is to analysis repeatability among reference metrics with FSOCT dataset. Therefore, before doing that we want to reduce dependent on dataset with confounding factors, e.g., SSI, age, axial length and gender. We performed unpaired T-test for age, SSI and axial length, and Fisher's exact test for gender to find association. The age and SSI have significantly associated ($P<0.05$) with the NFL reflectance metrics. However, in the subsequent stages of the analysis, we refrained from adjusting NFL reflectance using SSI as it had diagnostic information related to NFL intensity. Instead, we adjusted the normalized NFL reflectance metrics using age only.

### 2.3.5 FSOCT for repeatability, reproducibility and diagnostics

In the FSOCT dataset, multiple scans were performed for each eye. To evaluate within-visit repeatability and between-visit reproducibility, we divided the FSOCT data into two groups based on scan and visit. Two scans were selected for each participant on the same day for within-visit analysis. The duration between two visits was 5 to 14 months for between-visit analysis. The eye and scan were excluded if doesn't meet the within-visit and between-visit criteria.

The pooled standard deviation (pooled SD) has been estimated among all the participant for repeatability and reproducibility analysis. Moreover, the intra-class correlation coefficient (ICC) is also measured to assess the consistency or agreement among all of the subjects. The ICC has been implemented using linear mixed model to consider random effect. We estimated pooled SD and ICC for each of the references (NFL/PPEC Max, NFL/PPEC Mean, NFL/Post-NFL-Mean, NFL/Post-NFL-Bright, NFL/AC) for evaluation.

It could be noted lower pooled SD and higher ICC provide better repeatability and reproducibility that ultimately means higher diagnostic accuracy. Therefore, we have estimated AROC in R for diagnostic accuracy for each of the references.

# 3 Results

## 3.1 Neutral Density Filter Experiment

A total of 20 healthy participants were scanned at a 6x6 mm disc scan using Solix SD-OCT for the NDF experiment. With 7 optical densities levels, no filter and NDF with 0.1~0.6 optical density, we gathered a total of 140 scans from 20 participants. The average age and axial length were 35.4±8.0 years and 25.0±1.3 mm, respectively. Twenty percent participants were female.

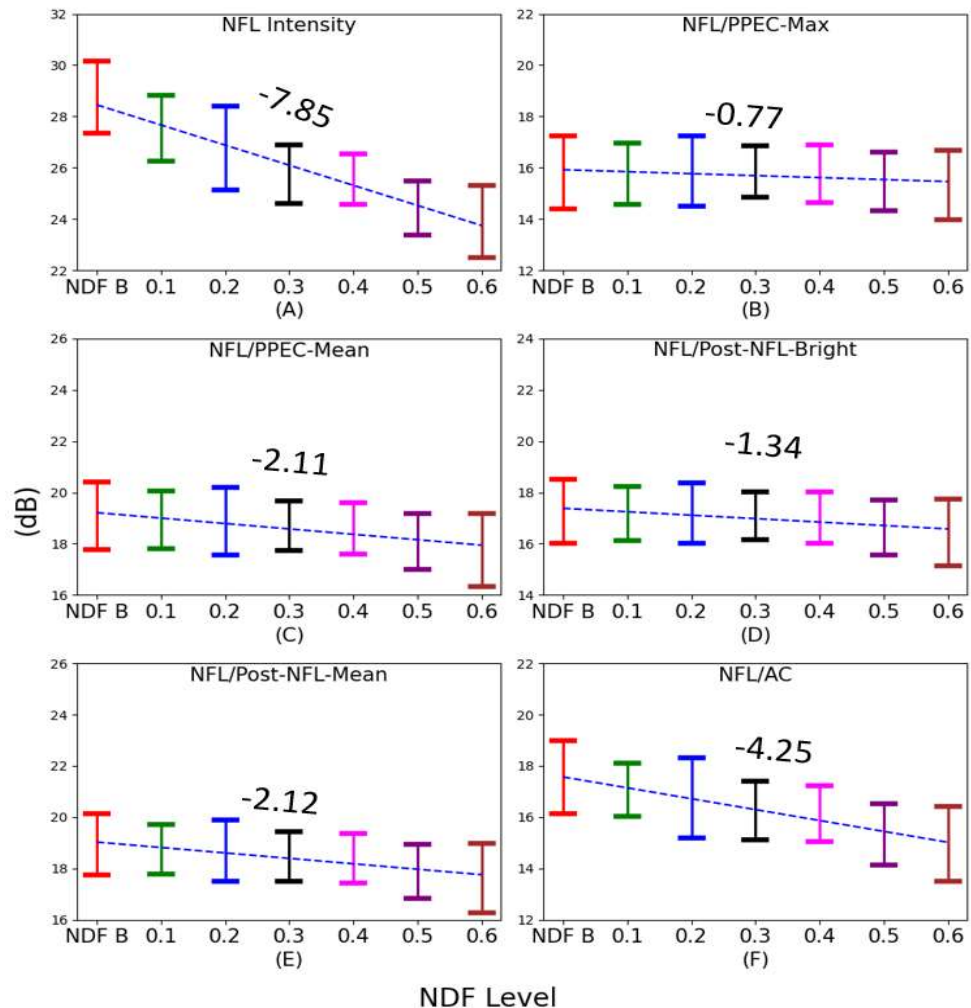

**Figure 2.** The association of nerve fiber layer (NFL) reflectance (unit: dB) normalized by difference references and neutral density filter (NDF) level unit: optical density). (A) NFL reflectance with reference; NFL reflectance normalized (B) by the maximum in pigment epithelium complex (PPEC), NFL/PPEC-Max; (C) by the average of PPEC, NFL/PPEC-Mean; (D) by the average the top 50% pixels in layers between ganglion cell layer and Bruch membrane, NFL/Post-NFL-Bright; (E) by the average in layers between ganglion cell layer and Bruch membrane , NFL/Post-NFL-Mean; (F) summation of attenuation coefficient in NFL layers, NFL/AC. Slopes were based on linear regression with a mixed model. NDF B=NDF baseline

We observed slopes slightly changed when the NDF level>=0.4 in figure 2. So we evaluated the difference of slope between two segments using a broken stick model with a break-point at NDF level =0.4 optical density (Table 1). The difference of slopes between two segments were significant (p<0.05) for normalized NFL reflectance using different sub-retinal layers, but not significant for the NFL reflectance without reference or the summation of NFL attenuation coefficients. The signal strength index (SSI) of scans was is 48.55±5.57 at NDF level=0.4. This indicated we should still choose a slight larger than 49 as a cutoff of image quality for NFL reflectance analysis after we used a reference layer to compensate the shadowing artefacts.

**Table 1.** Difference of the association of normalized nerve fiber layer (NFL) reflectance and neutral density filter (NDF) level between two segments

| Segment | NFL/PPEC Max | NFL/PPEC Mean | NFL/Post-NFL | NFL/Post-NFL-Bright | NFL/AC | NFL intensity |
|---|---|---|---|---|---|---|
| NDF_level <=4 | -0.01 | -1.46* | -1.67* | -0.76 | -4.42* | -9.84* |
| NDF level>4 | -0.67* | -2.03* | -2.06* | -1.26 | -4.27* | -8.10* |

NFL=nerve fiber layer; PPEC-Max: pigment epithelium complex-maximum intensity; NFL/PPEC-Max=NFL reflectance normalized by PPEC max; PPEC-Mean= PPEC-Mean intensity; NFL/PPEC-Mean= NFL reflectance normalized by PPEC-mean; *P-value <0.05

**Table 2.** Demography of FSOCT data

|  | *Normal* | *Glaucoma* | *PPG* | *PG* |
|---|---|---|---|---|
| *Participant* | 22 | 55 | 28 | 27 |
| *Eye* | 22 | 55 | 28 | 27 |
| *Age (Years)* | 59.7±8.7 | 64.5±10.0 | 64.5±9.9 | 64.5±10.2 |
| *Female (%)* | 68% | 58% | 64% | 51% |
| *Axial length* | 23.85±0.80 | 24.56±0.88 | 24.57±0.87 | 24.56±0.91 |
| *CCT (um)* | 546.7±34.4 | 538.2±36.8 | 532.3±37.2 | 544.4±35.7 |
| *VFMD (dB)* | 0.16±0.94 | -2.93±4.49 | -1.30±2.34 | -4.61±5.48 |

CCT=Central cornea thickness; VFMD = Visual field mean deviation; FSOCT= Functional and structural optical coherence tomography

### 3.2 FSOCT Clinic Data Analysis

Following the rules in selecting eyes with two visits and qualified repeated scans in each visit, we selected 22 normal eyes of 22 participants and 55 glaucoma eyes of 55 participants from FSOCT study (Table 2). Out of glaucoma eyes, 28 eyes were PPG and 27 were PG. The average ages were 58±10 years and 64±10 years and for normal and glaucoma participants (p<0.05, T-test), respectively. Females were 59% among all eyes (p==0.45, fisher's exact test) according fisher. The axial lengths were 23.8±1.0 mm for normal participants and 24.6±1.1 for glaucoma eyes (p<0.05, T-test).

We evaluated the associations of NFL reflectance with gender, age, axial length and SSI for different normalized NFL reflectance using univariate linear regression. To keep the data independent from the normal eyes used in later process, we select the 20 normal eyes in NDF experiment (scans without NDF) and 19 normal eyes in FSOCT study which did have the second visits. For all normalized NFL reflectance metrics, only age and SSI were significantly associated with NFL reflectance (p=0.01 for age & p<0.001 for SSI). Although SSIs were significantly associated with normalized NFL reflectance metrics, we didn't adjust NFL reflectance using the SSI in later process since the SSI was related to NFL intensity and therefore contained diagnostics information. The later analysis was done with normalized NFL reflectance metrics adjusted by age only.

All normalized NFL reflectance metrics reduced with increasing glaucoma severity (Table 3). Normal eyes had significantly higher NFL reflectance than PPG, PG, or glaucomatous eyes (p<0.01). NFL/post-NFL-Mean and NFL/post-NFL-Bright had the best within-visit and between-visit repeatability (table 4 and 5). The pooled SD for this two metrics was 20-25% smaller than other metrics. But the difference are not significant (only 28/3 comparisons were significant). We also noticed a trend of increasing repeatability and reproducibility with more subretinal layers in the reference layer. However, this trend did not apply to the NFL/AC, which used all pixels below NFL for compensation but had similar ICC and pooled SD as the NFL reflectance normalized by PPEC-max. The low repeatability of NFL/AC may be due to that the signal noise ratio behind Bruch membrane is worse than SNR in sub-retinal layers in spectral domain OCT.

The diagnostic accuracy of all normalized NFL reflectance were similar (AROC=0.82~0.87, and Sensitivity at 95% specificity: 47.88% to 53.33%). However, Post-NFL-Bright provide the best diagnostic accuracy, where the NFL/PPEC-Max is significantly lower than NFL/Post-NFL-Bright (p<0.05).

**Table 3.** Population Mean and standard deviation of normalized nerve fiber layer (NFL) reflectance

*Normalized NFL reflectance (average± standard deviation)*

|  | NFL/PPEC- Max | NFL/PPEC- Mean | NFL/Post- NFL | NFL/Post- NFL-Bright | NFL/AC |
|---|---|---|---|---|---|
| *Normal* | 17.32±0.79 | 18.35±0.77 | 18.04±0.74 | 17.52±0.80 | 16.80±0.66 |
| *PPG* | 16.13±1.33 | 17.14±1.31 | 16.76±1.26 | 16.13±1.32 | 15.61±1.20 |
| *PG* | 14.95±1.61 | 15.99±1.54 | 15.57±1.45 | 14.95±1.41 | 14.59±1.49 |
| *Glaucoma* | 15.55±1.58 | 16.57±1.53 | 16.17±1.47 | 15.55±1.78 | 15.11±1.43 |

PPG= Preperimetric glaucoma; PG= Perimetric glaucoma

**Table 4.** Repeatability and Reproducibility of normalized nerve fiber layer (NFL) reflectance and Intra-class Correlation for normal

| References | Pooled standard deviation (DB) | | Intra-Class correlation | |
|---|---|---|---|---|
|  | Within-visit | Between-visit | Within-visit | Between-visit |
| NFL/ PPEC-Max | 0.55 | 0.89 | 0.52 | 0.34 |
| NFL/ PPEC-Mean | 0.52 | 0.83 | 0.54 | 0.33 |
| NFL/Post-NFL-Bright | 0.43 | 0.79 | 0.64 | 0.37 |
| NFL/Post-NFL-Mean | 0.43 | 0.75 | 0.62 | 0.38 |
| NFL/AC | 0.53 | 0.79 | 0.45 | 0.30 |

**Table 5.** Repeatability and Reproducibility of normalized nerve fiber layer (NFL) reflectance and Intra-class Correlation for glaucoma

| REFERENCES | POOLED STANDARD DEVIATION (DB) | | INTRA-CLASS CORRELATION | |
|---|---|---|---|---|
| | Within-visit | Between-visit | Within-visit | Between-visit |
| NFL/PPEC MAX | 0.41 | 0.60 | 0.88 | 0.77 |
| NFL/ PPEC MEAN | 0.39 | 0.58 | 0.89 | 0.77 |
| NFL/POST-*BRIGHT* | 0.33 | 0.48 | 0.91 | 0.83 |
| NFL/POST-NFL-*MEAN* | 0.31 | 0.46 | 0.92 | 0.84 |
| NFL/AC | 0.40 | 0.61 | 0.87 | 0.73 |

**Table 6.** Diagnostic power among metrics

| Metrics | AROC (95% CI) | Sensitivity at 95% Specificity (95% CI) |
|---|---|---|
| NFL/PPEC-Max | 0.82 (0.768 to 0.856) | 53.33% (35.15 to 64.85%) |
| NFL/PPEC- mean | 0.84 (0.784 to 0.889) | 53.33% (32.73 to 66.06%) |
| NFL/Post-NFL-Bright | 0.87 (0.826 to 0.917) | 60.0% (42.42 to 72.33%) |
| NFL/Post-NFL | 0.86 (0.818 to 0.911) | 62.42% (41.21 to 73.33%) |
| NFL/AC | 0.84 (0.787 to 0.891) | 47.88% (36.97 to 71.52%) |

## 4 Discussion

As per the experimental results, NFL/PPEC Max is better in each NDF level followed by the Post-NFL-Bright, PPEC/Mean, Post-NFL-Mean and NFL/AC. We also observed more layers give better repeatability since it can reduce variation in the same person. For instance, NFL/Post-NFL-Mean and NFL/Post-NFL-Bright were used more layers below NFL and hence provide better repeatability and reproducibility that ultimately increase diagnostic accuracy.

The NFL/AC provides the lowest result in both NDF and ICC even though it has more layers, it might be the layer below PPEC contains more noise which affects the NFL/AC.

All of the reference metrics provide quite similar diagnostics accuracy (0.82-0.87), where NFL/Post-NFL-Bright provide the best result while the remaining were very competitive. It could be noted that each reference except NFL/AC provides better compensation till optical density 0.3, which is equivalent to image quality level 49.

Although the Post-NFL-Mean band offers better repeatability in FSOCT, it compensates poorly at each NDF level. This could be attributed to the presence of a darker layer in this band. To address this issue, we utilized the Post-NFL band and selected the top 50% of pixels, creating a new band called Post-NFL-Bright. By doing so, we were able to reduce the dependence on global attenuation after PPEC-Max. Thus the Post-NFL-Bright is more reliable reference since it perform well in both FSOCT and NDF dataset.

## 5 Conclusions

NFL reflectance corrected by NFL/PPEC-Max and NFL/Post-NFL-Bright provides better compensation in NDF datasets, while the summation of attenuation coefficients gave worse results. NFL/Post-NFL-Mean and NFL/Post-NFL-Bright showed better repeatability and diagnostics accuracy in FSOCT datasets. Overall NFL/Post-NFL-Bright is more reliable since it works well in both NDF and FSOCT datasets.

**Author Contributions**

O.T. and D.H. designed the study. O.T. and D. H. wrote the manuscript and all coauthors critically commented and/or edited the manuscript. D.H. supervised the project. L.L. and Q.Y. did the manual grading. J.W. and Y.L. developed the Center for Ophthalmic Optics & Lasers-Angiography Reading Toolkit (COOL-ART) software. A.C, E, I and J.M. conduct the clinical study.

**Competing Interests Statement**

OHSU, Dr. Tan, Dr. Huang, and Dr. Jia have a significant financial interest in Optovue, Inc., a company that may have a commercial interest in the results of this research and technology. These potential conflicts of interest has been reviewed and managed by OHSU.

on Signal Strength and Peripapillary Retinal Nerve Fiber Layer Optical Coherence Tomography Measurements. Journal of Glaucoma 20(1):p 37-43, January 2011. | DOI: 10.1097/IJG.0b013e3181ccb93b